# Cutting Last Wires for Mobile Communications by Microwave Power Transfer

Kaibin Huang [1] and Xiangyun Zhou [2]

## Abstract

The advancements in *microwave power transfer* (MPT) over past decades have enabled wireless power transfer over long distances. The latest breakthroughs in wireless communication, namely massive MIMO, small cells and millimeter-wave communication, make wireless networks suitable platforms for implementing MPT. This can lead to the elimination of the "last wires" connecting mobile devices to the grid for recharging, thereby tackling a long-standing ICT grand challenge. Furthermore, the seamless integration between MPT and wireless communication opens a new area called *wirelessly powered communications* (WPC) where many new research directions arise e.g., simultaneous information-and-power transfer, WPC network architectures, and techniques for safe and efficient WPC. This article provides an introduction to WPC by describing the key features of WPC, shedding light on a set of frequently asked questions, and identifying the key design issues and discussing possible solutions.

## 1. Introduction

Past decades have seen the explosive growth of wireless communications. A sequence of breakthroughs such as MIMO, capacity achieving codes, millimeter-wave communications and small-cell networks have achieved gigabit speeds for wireless access. As wireless and wire access speeds are becoming comparable, mobile devices, including smartphones and tablet and laptop computers, have replaced desktop computers as the dominant platforms for Internet access. In contrast, the advancements in battery technologies have been much slower. The resultant short battery lives require mobile devices to be periodically tethered to the grid for battery recharging. The cables for recharging are the last barrier for the devices to attain true mobility and thus called "last wires" in this article. As mobile services have penetrated different fields of the modern society such as banking, health care and civil defense, the interruption due to dead batteries can cause issues far more severe than merely inconvenience, such as financial loss and threats to health and public safety. Moreover, the production of billions of non-recyclable chargers per year poses a serious environmental issue. The urgency of addressing these issues and the existence of many market opportunities have recently motivated both the industry and academia to direct huge effort and funding towards developing technologies for wireless power transfer. Breakthroughs in such technologies will solve the grand ICT challenge of cutting the "last wires".

The idea of wireless power transfer using radio waves was first conceived and experimented by Nicola Tesla in 1899. However, the area did not pick up till 1960s when microwave technologies rapidly advanced, opening an active research field called *microwave power transfer* (MPT) [1]. In particular, the availability of large-scale antenna arrays and high-power microwave generators enables beaming of high power in a desirable direction. Moreover, the invention of *rectifying antennas* (rectennas) renders energy-conversion loss to a level practically negligible. Many advanced MPT systems have been designed such as wirelessly powered airborne vehicles that require no refueling and solar power satellites [1]. However, the enormous antenna arrays (e.g., arrays with diameters of hundreds of meters) that are instrumental for efficient MPT in such systems are impractical for everyday-life applications. This issue together with safety concerns has been delaying the commercialization of MPT. On the contrary, without such issues, non-radiative technologies for wireless power transfer, namely inductive coupling and resonant coupling between two coils, have been standardized and widely commercialized in mobile devices, home appliances and electric vehicles (see e.g., [2]). However, such technologies have the drawbacks of extremely short transfer distances (typically no more than a meter) and lack of support for mobility.

---

[1] K. Huang is with the Dept. of Electrical and Electronic Engineering at The University of Hong Kong, Hong Kong. Email: huangkb@eee.hku.hk.

[2] X. Zhou is with the Research School of Engineering at the Australian National University, Canberra, Australia. Email: xiangyun.zhou@anu.edu.au.

With long propagation ranges and support of mobility and multicasting, MPT appears to be a promising candidate technology for cutting the "last wires" if the two main challenges, *high propagation loss* and *safety concerns*, can be overcome. Both relying on microwaves as transmission vehicles, MPT and wireless communication have interwound R&D histories, yielding many common techniques and theories e.g., beamforming and propagation. The similarity allows communication techniques and network designs to be applied to tackle the mentioned MPT challenges. Specifically, sophisticated signal processing techniques for channel estimation, power control and adaptive beamforming can be adopted to ensure safety in MPT. Moreover, the latest breakthroughs in wireless communications, namely small cells, transmission using large-scale antenna arrays and millimeter-wave communications will dramatically reduce transmission distances and enable sharp beamforming, which will suppress propagation loss and achieve high power-transfer efficiencies. The availability of enabling technologies suggests that the time has come for cutting the "last wires", opening up the new area of *wirelessly powered communications* (WPC) with many exciting new research opportunities and applications. This article will introduce WPC by discussing its key features, answering a set of frequently asked questions, and identifying the key design challenges.

## 2. Key Features of Wirelessly Powered Communications

In this section, we introduce several key features of WPC including the MPT, mobile architecture supporting energy harvesting and *simultaneous wireless information-and-power transfer* (SWIPT).

### 2.1 Microwave Power Transfer

*Power Beamforming*

Efficient MPT hinges on concentrating radiated power in the direction of a target mobile by forming a microwave beam. Such beamforming for the sole purpose of power transfer is called *power beamforming*. A beam can be formed using an antenna array (or aperture antenna). The elements of the array are arranged with separation no larger than a half wavelength so as to avoid "grating lobes" (multiple beams). Under this constraint, the beam sharpness increases with the array size (or equivalently the number of elements). Sharp beamforming and short propagation distances are the key conditions for efficient MPT. They can be achieved by two corresponding latest wireless-communication technologies, namely *large-scale antenna arrays* (with hundreds to thousands of antenna elements) and *small cells*, currently under extensive development and expected to be deployed in next-generation wireless networks [3].

With the array size fixed, the beam sharpness can be increased by scaling up the carrier frequency and correspondingly packing more antennas into the array. Traditional MPT without a dedicated spectrum uses the carrier frequency of either 2.4 GHz or 5.8 GHz in the ISM band [1]. However, with the rapid advancement of millimeter-wave communication, the MPT embedded in WPC can be operated in the 60-GHz bandwidth in the near future. Such high frequencies enable ultra-sharp beamforming even when the array size is small, leading to dramatically improved power-transfer efficiencies.

*Power-Transfer Channel and Beam Efficiency*

Rich scattering is typical in a wireless communication channel and can be combined with transmit and receive antenna arrays to support multiple parallel data streams without requiring additional bandwidth. In contrast, free-space propagation is essential for power beamforming, since a scatterer can disperse power beam and cause the transfer efficiency to drop dramatically. Thus, a power-transfer channel refers to one over free space. The propagation distance ranges from the *near field*, where the distance is comparable with the transmit array dimension, to the *far field*. The factors determining the propagation loss include: 1) the apertures of the transmit and receive arrays, denoted as $A_t$ and $A_r$, respectively, 2) the wavelength $\lambda$ and 3) the propagation distances $d$ as elaborated in the sequel.

The end-to-end MPT efficiency is equal to the product of three efficiencies: 1) DC-to-RF power-conversion efficiency, 2) *beam efficiency* defined as the ratio between the received and radiated powers, and 3) RF-to-DC power-conversion efficiency. The state-of-the-art microwave generators and rectennas can achieve close-to-one values (e.g., 80%) for the first and third efficiencies, respectively [1]. Therefore, the beam efficiency is the bottleneck for efficient MPT over long distances.

Consider a pair of transmit/receive circular aperture antennas (that can be replaced by arrays with the same apertures) facing each other over a power-transfer channel. For this scenario, the beam efficiency can be accurately approximated as [1]

$$\text{Beam Efficiency} = 1 - e^{-\beta} \tag{1}$$

where $\beta$ is given as

$$\beta = \frac{A_t A_r}{(\lambda d)^2}. \tag{2}$$

Note that (2) is equivalent to the Friis equation for far-field transmission. The propagation as described in (1) covers both the near and far fields. For the far field where $\beta$ is small (large $d$), the propagation loss is approximately equal to $\beta$, thus following the Friis transmission equation. For the near field where $d$ is small and hence $\beta$ is large, the beam efficiency is close to one. The beam efficiency is plotted in Fig. 1 as a function of the ratio between the transfer distance and the receiver antenna radius, where the carrier frequency is 2.4 GHz. Next, (1) suggests the trade-offs between the MPT parameters $A_r, A_t, \lambda$ and $d$. In particular, for a given beam efficiency, doubling the transmit-array radius or the carrier frequency doubles the transfer distance or supports recharging of smaller (half-size) mobiles. For instance, scaling up the frequency from 2.4 GHz to 60 GHz in the millimeter band increases the power-transfer distance by 25 times. This, however, requires the numbers of transmit/receive antennas to grow by the square of this factor if the aperture antennas are replaced by antenna arrays.

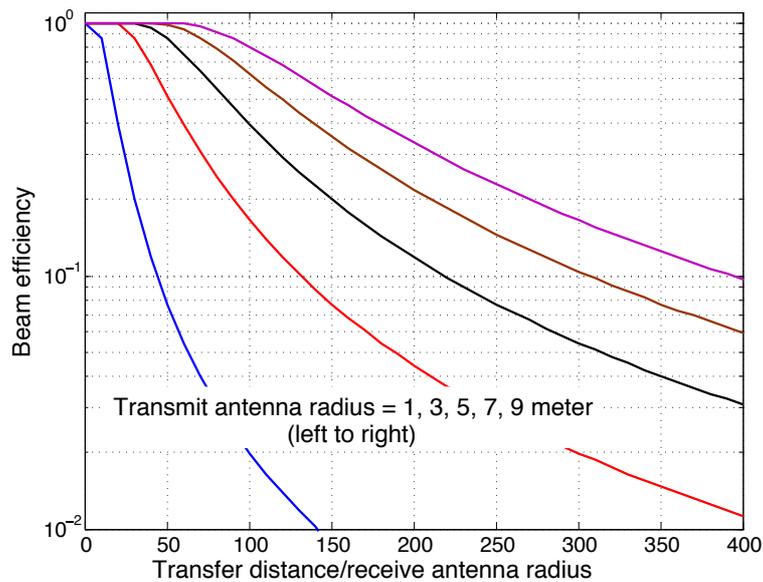

*Figure 1. Beam efficiency for MPT versus the ratio between the transfer distance and the receiver antenna radius for a carrier frequency of 2.4 GHz.*

## 2.2 Mobile Architecture for WPC

A traditional mobile device comprises an information transceiver powered by a rechargeable battery. For WPC, a RF energy harvester is included in the mobile device for harvesting energy from the incident microwave signal to power the transceiver as shown in Fig. 2. The design of an energy harvester is rather simple and consists of a rectifying circuit for converting the RF signal at the antenna output to DC power that is stored using a rechargeable battery or a super capacitor. The most popular and efficient design of the RF energy harvester uses a rectenna which integrates a single antenna and a rectifying circuit.

The functionality of the antennas used by the information transceiver and energy harvester is different. The array attached to the transceiver enables multi-antenna communication and array processing (e.g., receive beamforming and interference nulling). Therefore, it is desirable to have as many (small) antenna elements as possible. On the other hand, the rectenna requires to capture as much incident power as possible, hence, the rectenna design aims for the largest possible antenna aperture. This is an important tradeoff for designing a WPC receiver under a form-factor constraint.

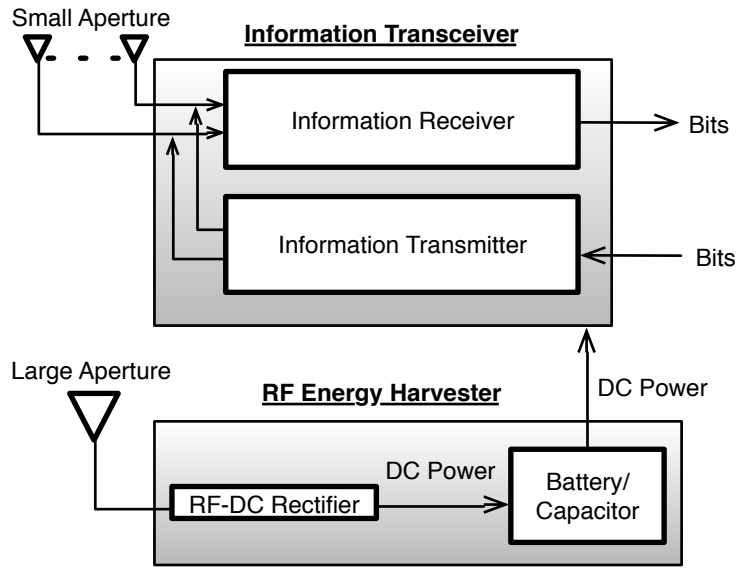

*Figure 2. Mobile architecture for WPC.*

## 2.3 Simultaneous Wireless Information-and-Power Transfer

Since the information transceiver and energy harvester have separate antennas and circuits, the mobile architecture supports SWIPT [4]. As illustrated in Fig. 3, there exist three designs of SWIPT system, namely *integrated SWIPT*, *closed-loop SWIPT*, and *decoupled SWIPT*, which are described as follows. Integrated SWIPT in Fig. 3(a) is the simplest design where power and information are extracted by the mobile from the same modulated microwave transmitted by a base station (BS). For this design, *information transfer* (IT) and *power transfer* (PT) distances are constrained to be equal. Closed-loop SWIPT in Fig. 3(b) consists of *downlink PT* and *uplink IT*. The signal power received at the BS originates from the BS radiated power and its closed-loop propagation (downlink+uplink) incurs double attenuation [5]. Thus closed-loop SWIPT only supports very short ranges and is unsuitable for cell-edge mobiles. Last, a decoupled-SWIPT system in Fig. 3(c) builds on the traditional communication system to include an additional special station, called *power beacon* (PB), dedicated for MPT to mobiles [6]. PT and IT are orthogonalized by using different frequency bands or time slots to avoid interference, giving the name decoupled SWIPT. Unlike BSs, PBs require no backhaul links and the resultant low cost allows dense deployment of PBs to enable efficient MPT.

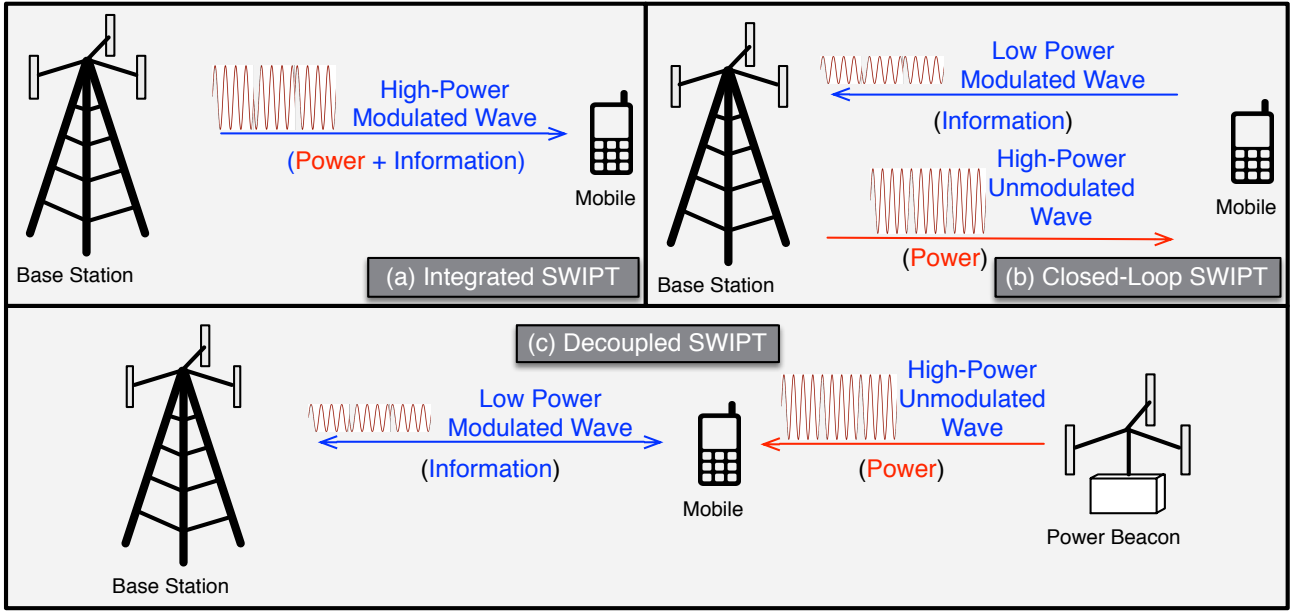

*Figure 3. Three system configurations for SWIPT: (a) integrated SWIPT, (b) closed-loop SWIPT and (c) decoupled SWIPT.*

## 3. Wirelessly Powered Communications: Frequently Asked Questions

Current research focuses on developing the WPC theory based on abstracted system models. Surprisingly, many frequently asked questions (FAQs) on e.g., the practicality and safety of WPC, remain unanswered. In this section, an attempt is made to shed light on some FAQs.

### 3.1 How far can a mobile device be wirelessly powered?

BSs can support communication ranges up to tens of kilometers. This can lead to the misconception that BSs/PBs can also power mobiles at comparable distances since both IT and PT use microwaves as vehicles. It is important to understand that the efficiency of PT depends on the received signal power while the reliability of IT is determined by the receive signal-to-noise ratio (SNR). Since the noise power is extremely low (e.g., -120 dBm), the received signal power for IT falls in the range of -100 dBm to -50 dBm, which is many orders of magnitude lower than the power consumption of mobile devices. Thus one should expect drastically shorter ranges for PT than those for IT. For comparison, the typical values for received signal power and the power consumption of popular mobile devices are listed as follows.

- **Wireless signals**: -120 to -50 dBm;
- **ZigBee devices or sensors**: 1 to 100 mW;
- **Smartphones**: 19 mW to 1.3 W;
- **Tablet computers**: 1 W to 11 W;
- **Laptop computers**: 19 W to 52 W.

One can see that the typical power consumption of mobile devices ranges from milliwatts for sensors or ZigBee devices to tens of watts for laptop computers, which are about 50-100 dB higher than the range of wireless signal power. In addition, the sensitivity level of a typical energy harvester is in the order of -10 dBm and below this level little energy can be harvested.

To get a concrete answer to the current question, the PT distances can be computed using practical settings. Consider the scenario where a PB wirelessly powers a mobile device where transmit and receive antenna apertures are assumed to be circular. The power-transfer ranges are computed numerically based on the beam-efficiency equation in (1) and the results are plotted in Fig. 4 for different transmitted powers. The specifications for the numerical computation are summarized in the caption of Fig. 4. It is surprising that the PT ranges for small to medium devices (ZigBee/sensors, smartphones, tablets) are very similar. The reason is that a larger device can harvest more power (with a larger antenna aperture) that compensates for the increase in power consumption. One can observe that a PB transmitting tens of watts can power sensors, smart phones and tablets at a distance of around 10 meters. Interestingly, this distance matches the smartphone recharging range of the MPT based charging station developed by a new startup called COTA [7]. The relative short PT distances in Fig. 4 suggest that PBs should be equipped with large-scale antenna arrays and installed in an indoor environment or with high density.

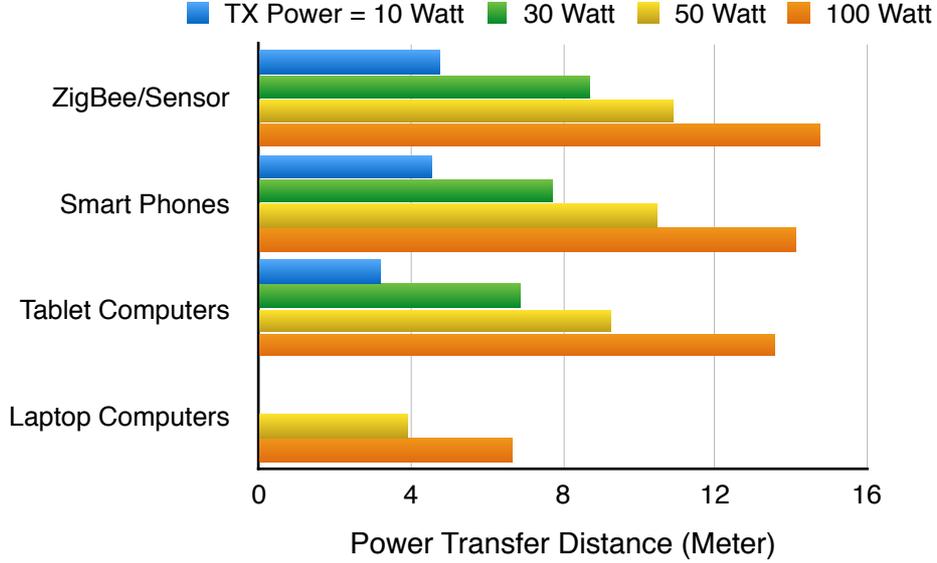

*Fig. 4. Power-transfer ranges for different types of mobile devices. The transmitter antenna array has about 260 elements separated by half wavelength and arranged on a disk, having an aperture of 3 m in radius. The carrier frequency is 2.5 GHz. The RF-to-DC conversion efficiency is 70%. The power consumption and antenna radiuses for different types of mobile devices are assumed to be (50 mW, 1 cm) for ZigBee devices, (0.5 Watt, 3 cm) for smart phones, (5 Watt, 9 cm) for tablet computers and (25 Watt, 11 cm) for laptop computers.*

To summarize, under practical constraints, the MPT distance is in the range of 3-15 meters for typical mobile devices depending on the radiated power. The MPT ranges for ZigBee devices, sensors and tablets are similar and abut twice of that for laptop computers. Such distances are still substantially shorter than cell radiuses of 5G small-cell networks which are 50-100 meters [3].

## 3.2 Is WPC safe?

By sharp power beamforming, the power density can be high along the path from a PB to a target mobile. This can potentially cause health hazards to a human body accidentally intercepting the path. According to international safety standards set by authorities such as FCC and ICNIRP (Table 3 in [8]), a person should not be exposed to microwave radiation with an average power density higher than 10 W/m$^2$ over a half-hour time window. Note that the wavefront area of a beam grows with the propagation distance and thus the power density of the beam decreases accordingly. We can define the *unsafe beam-interception distance* (UBID) as the maximum propagation distance where the beam power density exceeds the mentioned exposure limit. Assuming that a beam contains 90% of the total radiated power, the beam efficiency in (1) allows us to compute the UBIDs for different practical configurations as follows.

- UBDI = **0.63 meters** for radiated power *P* = 50 watts if the transmission is **omnidirectional**;

- UBDI = **14.4 metes** for *P* = 10 watts and beamed transmission with an antenna aperture = 3 m$^2$;

- UBDI = **32 meters** for *P* = 50 watts and beamed transmission with an antenna aperture = 3 m$^2$.

For omnidirectional transmission, the UBDI is found to less than a meter and hence there is no safety concern in practice. On the contrary, for power beamforming, one can see from the numbers that if a person is within 15 meters from the PB, he/she should be careful not to stand in-between the PB and the mobile for too long. The unsafe distance is higher for larger radiated power or a sharper beam (a larger antenna aperture). The PT distances in Fig. 4 are observed to be smaller than the above UBDI values, suggesting potential safety issues for WPC. However, the limits on microwave exposure as set by different authorities are average values over a long time window (0.5 hour). This is more than sufficient for an intelligent WPC system to adapt its transmission power (within e.g., milliseconds) to ensure safety.

In summary, though it is unsafe for a person to be illuminated by a power beam within the PT range for too long, intelligent beam-control techniques can be designed to ensure safety in WPC (see Section 4.1 for details).

### 3.3 Is SWIPT practical?

A main motivation for implementing integrated and closed-loop SWIPT (see Fig. 3) is that SWIPT can be realized by upgrading existing BSs (including WiFi access points) without changing the network architecture. However, constrained by the short PT ranges, BSs can perform SWIPT only to a small fraction of nearby mobiles. Supporting full network coverage for SWIPT requires the deployment of much denser BSs with extreme cell radiuses of 10-15 meters, which can result in enormous cost and is thus impractical. On the other hand, the decoupled SWIPT design (see Fig. 3) provides a more practical solution for network-wise SWIPT coverage due to the low deployment cost of backhaul-less PBs.

To provide an answer for the current question, a relatively practical solution of providing network coverage for SWIPT is its implementation based on decoupled SWIPT that exploits low-cost PBs for dense deployment for cutting "last wires".

### 3.4 Does MPT interfere with wireless communications?

MPT can interfere with wireless communications in several direct or indirect ways. First, for decoupled SWIPT, the information-carrying signal can be deeply buried in the power-carrying signal if they are simultaneously received by a mobile due to their orders of magnitude difference in power as mentioned earlier. Therefore, IT and PT must be sufficiently separated in frequency, e.g., a bandwidth should be reserved for the sole purpose of MPT. Second, even given sufficient separation in frequency, the power-carrying signal must be suppressed right at the antenna outputs of the mobile's information receiver rather than in the digital domain. Otherwise, the extremely strong signal can saturate the amplifier and ADC and thereby cause distortion and excessive quantization noise to the simultaneous information-carrying signal. Last, the circuit non-linearity of microwave generators can cause strong harmonics of the carrier that interfere with wireless communications.

In summary, MPT can interfere with wireless communications in practice unless countermeasures are implemented in the system design.

### 3.5 Is it possible to power mobile devices by RF energy scavenging?

Cellular BSs and WiFi access points are ubiquitous in the urban environment. Their transmissions result in the existence of RF energy in the ambient environment. Their high frequencies (e.g., 2.4 GHz for WiFi) require resonant antennas of relatively small sizes (10-50 $cm^2$) for RF energy harvesting. Scavenging such energy for powering mobile devices is a green approach since it does not require installation of additional power sources. The amount of power that can be generated by energy scavenging depends on the power density. Some available measurement results are summarized in Table 1 [9, 10]. One can see that the maximum power density in the order of $mW/m^2$. Therefore, a mobile device (e.g., a smartphone) of a typical size smaller than 100 $cm^2$ can harvest peak power of tens of μW with the average in the order of μW. This gives the following answer. RF energy scavenging is sufficient only for powering small sensors with sporadic activities. Wirelessly powering larger devices has to rely on dedicated PBs.

*Table 1. Measured Power Densities of RF Signals in the Ambient Environment*

| Spectrum | Environment | Power Density (mW per sq. m) |
|---|---|---|
| **GSM (935 - 960 MHz)** | inner city, outdoor, on ground | $10^{-3} - 10^{-1}$ |
| | inner city, indoor, close to window | $10^{-2} - 10^{-1}$ |
| **GSM (1805 - 1880 MHz)** | 50 meters from base stations | $5 \times 10^{-3} - 5$ |
| | 200 meters from base stations | $10^{-3} - 0.5$ |
| | 500 meters from base stations | $5 \times 10^{-4} - 5 \times 10^{-2}$ |
| **WiFi** | within 8 meters from access points | $10^{-3} - 5 \times 10^{-2}$ |
| | 12 meters from access points | $10^{-4} - 5 \times 10^{-4}$ |

## 4. Wirelessly Powered Communications: Designs and Challenges

In this section, we discuss the key design considerations and challenges for WPC.

### 4.1 Efficient and Safe WPC

*Pilot Signal Design for Retrodirective Beam Control*

Power beamforming for MPT typically uses a phase array and the technique called retrodirective beam control that automatically steers a beam in the reverse direction of the incident pilot signal sent by the mobile by exploiting channel reciprocity. The simple procedure for retrodirective beam control is as follows:

1) The receiver transmits a pilot signal;
2) The transmitter computes the phase shift of the output of each antenna by comparing it with a local reference signal;
3) The phase shift for each antenna is conjugated and applied to the phase shifter for transmission.

An important aspect of implementing retrodirective beam control for WPC is the design of pilot sequences. They should be designed to initiate MPT for multiple mobiles at the same time. However, mobiles in different cells may transmit non-orthogonal or identical pilot sequences, resulting in *pilot contamination* which lays a fundamental limit for the performance of a cellular network equipped with large-scale antenna arrays [11]. In WPC networks, pilot contamination not only degrades the IT performance but also decreases PT efficiency. To be specific, receiving multiple pilot signals can cause the retrodirective beamformer for MPT to auto-reflect multiple beams towards both the intended and unintended mobiles, which reduces the beam efficiency to the former and, more importantly, causes safety threats to people in unintended directions. Tackling pilot contamination continues to be a key challenge in designing WPC networks.

In addition, the power of pilot signals and duty cycle of pilot transmission should be designed to optimize the tradeoffs between the PT efficiency, training overhead and mobile energy consumption.

*Safety Measures*

Measures for ensuring safe PT is a unique and important aspect of designing WPC systems. The retrodirective beamformer has the safety feature that it can automatically de-phase a beam when it is intercepted by an object such as a human body. This measure, however, does not protect people who are near a target mobile but do not intercept the beam. Thus, additional safety measures are required. For example, guard zones can be created around not only the PBs but also the mobiles using technologies such as microwave life detection [12]. Other technologies such as surveillance cameras, radar tracking and network localization can be deployed for accurate human detection and thereby further enhancing the safety in WPC.

*Efficient and Safe Power Transfer Using Multiple Coordinated Power Beacons*

In a WPC network deploying dense PBs, a single mobile can be powered by multiple coordinated PBs based on the idea proposed in [13]. The PBs surrounding the target mobile form multiple incoming power beams from omni-directions that are coherently combined at the mobile location due to beacon coordination. PT using coordinated beacons is safe for two reasons. First, many incoming beams from different directions enable the detection of human presence at practically any arbitrary location near the target mobile, thereby overcoming the drawback of stand-alone PBs. Next, incident power beams from omni-directions have the combined effect of concentrating transmitted power at the mobile and very low power density at other locations. Moreover, multi-beams improves the chance of finding lines-of-sight for efficient MPT. As the combined result, MPT using coordinated beacons also improves the PT efficiency.

## 4.2 WPC Network Architecture

A WPC network is designed to deliver two types of services, high-speed wireless access and MPT, to mobile devices and its performance is measured by the coverages of both services. As illustrated in Fig. 5, the WPC network comprises BSs, PBs and mobile devices. The main role of BSs is to provide network-wise wireless-access coverage while they can also support SWIPT to nearby mobiles. Full MPT coverage is achieved by deploying dense PBs for supporting MPT to mobiles. Mobiles can be separated into receiving and transmitting mobiles. It is more challenging to wirelessly power transmitting mobiles since they require additional power for transmission while both types of mobiles consume circuit power.

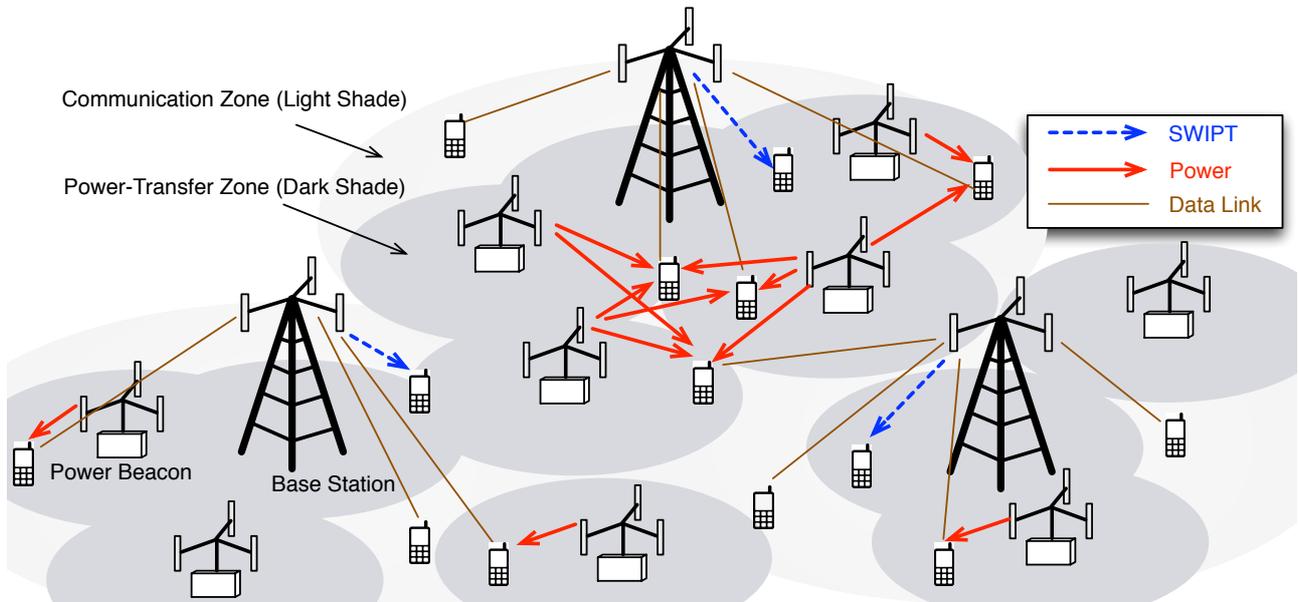

*Figure 5. WPC network architecture.*

For designing a WPC network, one of the first challenges is to understand the required densities of BSs and PBs for providing network coverage for both wireless access and wireless power. Recently, the tradeoff between these densities was quantified in [6] by modeling the WPC network using stochastic geometry and under reliability constraints on the network services. This tractable approach can be extended to design WPC networks with more complex architectures such as heterogeneous BSs/PBs. Apart from the fixed deployment of PBs, mobile PBs can be also deployed to support wider coverage with fewer PBs or more efficient MPT by shortening the transfer distances. One challenge there is to design the optimal routing for each mobile PB.

## 4.3 WPC Protocols and Techniques

Compared with traditional wireless networks, the addition of PBs and the interaction between IT and PT enrich the WPC network architecture and the operational modes of network nodes. As a result, traditional communication protocols and techniques must be thoroughly redesigned to enable efficient WPC. Several key challenges are identified as follows, which point to promising research directions.

- **Cognitive WPC**: The principle of cognitive radios can be applied to design cognitive WPC systems to enable seamless integration of PT and IT and accommodate passive secondary nodes (see e.g., [14]). In particular, a cognitive PB can sense the spectrum and choose a proper subset of frequency sub-channels for MPT to avoid interfering with IT and at the same time reduce power-spectrum density to meet the safety requirements set by authorities.

- **Cooperative PB/BS clustering**: Grouping PBs/BSs for cooperation enhances the PT efficiency (refer to Section 4.1) besides mitigates interference in IT. However, PB/BS clustering is much more complex than that for traditional multi-cell cooperation due to many new factors for consideration including multi-user beam efficiencies, BS modes (SWIPT or IT only) and wireless signaling overhead between backhaul-less PBs.

- **Relay Assisted WPC:** In WPC systems with relatively sparse BSs/PBs, mobiles far away from them receive double penalties: lower PT efficiencies but larger power required for uplink transmission. Thus, it is critical to address the issue of fairness in designing such systems. An alternative cost-effect approach apart from deploying dense PBs is to motivate mobiles to cooperate by relaying information/power for each others or deploying dedicated passive relay stations (see e.g., [15]). This opens many new research issues on relay-assisted WPC ranging from the signal processing methods, scheduling and MAC protocols, and network performance.

- **Joint scheduling and resource allocation of PT and IT**: For WPC networks of low-complexity and low-power devices, the communication protocols are often simple and predetermined. In contrast, for scenarios where the mobiles are able to handle complex algorithms, the optimal solution is to design and deploy intelligent communication and resource allocation algorithms for mobiles, PBs and BSs that are adapted to the dynamic states of mobile energy storage, data queues, channels and beam efficiencies.

## 5. Towards Truly Mobile Communications

Cutting the "last wires" of mobile devices will endow them the long desired immortality, which will bring users convenience, strengthen the reliability of widespread mobile services and create a huge range of market opportunities. This task is far more than straightforward implementation of the MPT technology but requires a seamless integration between information and power transfers. As a result, many new research challenges arise including designing network architectures for enabling SWIPT, achieving highly efficient and safe MPT to mobile devices, and revamping traditional communication techniques, such as cooperation, cognitive radios and adaptive transceivers, to integrate power transfer into communication networks. This leads to a newly emerged area called wirelessly powered communications. It is through the advancements in this area and relevant areas such as energy scavenging, batteries and low-power electronics that the tens of billions of devices to be deployed in the coming decade will be free from the "last wires" and attain true mobility.

## Acknowledgement

We thank Dr. Rahul Vaze and Dr. Salman Durrani for their comments that have significantly improved the presentation of this article.

**Kaibin Huang** (S'05–M'08-SM'13) received the Ph.D. degree from The University of Texas at Austin in electrical engineering. Since Jan. 2014, he has been an assistant professor in the Dept. of EEE at the University of Hong Kong. He is a guest editor for the IEEE JSAC, an editor for the IEEE Transactions on Wireless Communications and IEEE Wireless Communications Letters. Dr. Huang received a Best Paper Award from IEEE GLOBECOM 2006. His research interests focus on the analysis and design of wireless networks using stochastic geometry and multi-antenna techniques.

**Xiangyun Zhou** is a senior lecturer at the Australian National University (ANU). He received the Ph.D. degree in telecommunications engineering from ANU in 2010. His research interests are in the fields of communication theory and wireless networks. He serves on the editorial board of IEEE Transactions on Wireless Communications and IEEE Communications Letters. He is a recipient of the Best Paper Award at the 2011 IEEE International Conference on Communications.